\documentstyle[12pt,epsfig]{article}
\begin{document}
\baselineskip=23pt
 ~~~~~~~~~~~~~~~~~~~~~~~~~~~~~~~~~~~~~~~~~~~~~~~~~~~~~~~~~~~~~~~~~~~~~~~~~~BIHEP-TH-2002-21

\vspace{1.2cm}

\begin{center}
{\Large \bf Thermodynamics of a deformed Bose gas}

\bigskip

Zhe Chang\footnote{changz@mail.ihep.ac.cn} and Shao-Xia Chen\footnote{ruxanna@mail.ihep.ac.cn}\\
{\em Institute of High Energy Physics,
Chinese Academy of Sciences} \\
{\em P.O.Box 918(4), 100039 Beijing, China}
\end{center}

\vspace{1.2cm}

\begin{abstract}
By making use of the double-time Green function technique, we
study thermodynamics of a deformed Bose gas, which describes well
properties of density intensive photonic gas and radiation fields
of the early universe. General form of statistical distribution
function is obtained. We show explicitly the expression of the
distribution function in some limitation cases. The free energy,
equation of state, specific heat and other thermodynamic
properties of the deformed Bose gas are presented.

\end{abstract}
\vspace{1.2cm}
\vspace{1.2cm}
\section{Introduction}

There are two main motivations to study a deformed many-body
system. The first one comes from fundamental physics. It is
well-known that the noncommutative geometry plays a very important
role in unravelling the properties of the Planck scale  physics.
It has for a long time been suspected that the noncommutative
spacetime might be a realistic picture of how spacetime behaves
near the Planck scale\cite{a05}. In fact, the noncommutative
geometry naturally enters the theory of open string in a
background $B$-field\cite{a06}. In particular, the noncommutative
geometry makes the holography\cite{a09} ({\em e.g.} the AdS/CFT
correspondence) of a higher dimensional quantum system of gravity
and lower dimensional theory possible. A kind of special
regularization with exponentially increasing spacetime cutoff has
been introduced based on the noncommutative geometry. A very small
displacement of the noncommutative deformation parameter from its
classical value reduces sharply the entropy of quantum system of
gravity. An adequately adopted noncommutative deformation of
geometry makes the holography of a higher dimensional quantum
system of gravity and lower dimensional theory possible. It was
also discovered that simple limits of $M$ theory and superstring
theory lead directly to the noncommutative gauge field
theory\cite{a07,a08}. The noncommutative field theory has been
intensively studied in the past two decads\cite{a11,a12}. The most
realistic laboratory for testing the Planck scale physics is the
early universe and the black hole physics. From the standard
picture of cosmology, the universe started from a very hot, dense
phase about 15 billion years ago in the big bang. The very early
universe was opaque due to the constant interchange of energy
between matter and radiation. Several problems of standard big
bang cosmology, such as the horizon, homogeneity, entropy and
flatness problems, can be explained under the assumption that the
very early universe underwent a period of extremely rapid
inflation\cite{a13}, driven by the potential of some assumed
inflation fields. In particular, the inflation scenario is also
able to explain the observed perturbations in the cosmic microwave
background radiation\cite{a14,a15}, namely as originating
ultimately from quantum fluctuations of the inflation field.
However, in most of the current models of inflation, the physical
length of perturbations was much smaller than the Planck length at
the beginning of inflation\cite{a16,a17}. It is then of interest
to investigate whether the predictions of the spectrum of
cosmological perturbations are sensitive to the unknown
super-Planck-scale physics\cite{a18}-\cite{a21}. Recent
studies\cite{a22}-\cite{a29} by physicists and astrophysicists
show that without strange inflation fields but just the
noncommutative geometry and the deformed radiation dominated
universe subject to the noncommutative spacetime quantization
indeed give rise to inflation as the radiation temperature exceeds
the Planck temperature. Based on the noncommutative geometry and
the deformed radiation, one can even get an alternative
explanation for the horizon, homogeneity, entropy and flatness
problems in the standard cosmology.

Another motivation comes from quantum optics as well as molecular
and atomic spectroscopy. A fascinating development of quantum
optics in last two decades has taken place within two specific
areas, one concerning the study of nonclassical states of light,
and the other studying the interaction of light with matter. The
nonclassical states of light offer the possibility of the
so-called reduction of quantum noise below the standard limit,
determined by the Heisenberg's uncertainty relation. Study on
interaction of light with matter shows that the old Bohr quantum
jumps are real. Dynamics of  deformed photonic field interacting
with matter presented many interesting properties both in the two
important areas\cite{a30}-\cite{a34}. Great success has also been
made by making use of a deformed oscillator to study molecular
spectrum\cite{a35}-\cite{a37}.

In this paper, we study thermodynamics of a deformed Bose gas, and
in particular, of a deformed radiation field. On the topic, many
papers have been published in more than ten
years\cite{5}-\cite{18}. However, we still have not a general form
of statistical distribution function of the deformed Bose gas,
which should reduce to the Bose-Einstein distribution function for
an ideal Bose gas in the $q=1$ limit. In this situation, it is
difficult to investigate systematically the influence of the
deformation on thermodynamic characteristics of a Bose system,
which is interesting in fundamental physics and other subjects as
discussed above. For this purpose, by making use of the
double-time Green function technique, we obtain an algebra
equation, which should be satisfied by the distribution function
of a deformed Bose gas. The equation for distribution function can
not be solved exactly in general. The situation is similar to the
case of magnetic lattices. From this point of view, a deformed
Bose gas is really a strongly correlated system. In the case of
$q\rightarrow 1$, we give a useful expression of the distribution
function for a deformed Bose gas. Thermodynamic properties, such
as internal energy, specific heat, free energy and equation of
state, are discussed in detail.

The paper is organized as follows. In section 2, a deformed Bose
gas is discussed. We get the distribution function for a deformed
Bose gas by using the double-time Green function technique in
section 3. In section 4, we calculate some thermodynamic
quantities for a deformed photon gas and compare them with those
for the ideal Bose gas. Concluding remarks are given in section 5.

\section{A deformed Bose gas system}
The Lagrange density of a scalar field $\phi(x)$ with mass $m$ reads
\begin{equation}
{\cal L}(x)=\frac{\hbar^2}{2}\frac{\partial\phi}{\partial x_\mu} \frac{\partial\phi}{\partial x^\mu}-\frac{1}{2}m^2c^2\phi^2~.
\end{equation}
The Euler-Lagrange equation immediately leads to the Klein-Gordon
equation
\begin{equation}
\left(\hbar^2\frac{\partial}{\partial x_\mu}\frac{\partial}{\partial x^\mu}+m^2c^2\right)\phi(x)=0~.
\end{equation}
The canonically conjugate field is
\begin{equation}
\pi(x)=\frac{\partial{\cal L}}{\partial\dot{\phi}(x)}~,
\end{equation}
which leads to the Hamilton density
\begin{equation}
{\cal H}(x)=\frac{1}{2}\left(\frac{1}{\hbar^2}\pi^2(x)+(\hbar\nabla\phi(x))^2+m^2c^2\phi^2(x)\right)~.
\end{equation}
The Fourier decomposition of the scalar field is of the form,
\begin{equation}
\begin{array}{l}
\phi(x)=\displaystyle\int\frac{d^3k~c}{\sqrt{(2\pi)^32\hbar\omega_k}}\left(a(k)e^{-ikx}+a^\dagger(k)e^{ikx}\right)~,\\
\pi(x)=\displaystyle\int\frac{d^3k~i\hbar^2\omega_k}{\sqrt{(2\pi)^32\hbar\omega_k}}\left(-a(k)e^{-ikx}+a^\dagger(k)e^{ikx}\right)~,
\end{array}
\end{equation}
where we have used the notation
$\omega_k=\frac{c}{\hbar}\sqrt{(\hbar {\bf k})^2+m^2c^2}$. The
canonical quantization of the scalar field reads
\begin{equation}
[\hat{\phi}({\bf x},t),\hat{\pi}({\bf y},t)]=i\hbar c\delta^3({\bf x}-{\bf y})~.
\end{equation}
In terms of the Fourier modes $\hat{a}(k)$ and
$\hat{a}^\dagger(k)$, the commutation relations can be rewritten
into the form
\begin{equation}
[\hat{a}(k),\hat{a}^\dagger(k')]=\delta^3({\bf k}-{\bf k}')~.
\end{equation}
It is not difficult to express the Hamiltonian in terms of these
Fourier modes,
\begin{equation}\label{hamiltonian}
\hat{H}=\frac{1}{2}\int d^3k~\hbar\omega_k\left(\hat{a}(k)\hat{a}^\dagger(k)+\hat{a}^\dagger(k)\hat{a}(k)\right)~.
\end{equation}
In the quantum field theory, the vacuum state is defined as
follows
\begin{equation}
\hat{a}(k)\vert 0\rangle=0~.
\end{equation}
And the $N$-particle Fock space can be written down directly as
\begin{equation}
\vert n(k_1),n(k_2),\cdots,n(k_m)\rangle=\frac{\left(\hat{a}^\dagger(k_1)\right)^{n(k_1)}}{\sqrt{n(k_1)!}} \frac{\left(
\hat{a}^\dagger(k_2)\right)^{n(k_2)}}{\sqrt{n(k_2)!}}
\cdots \frac{\left(\hat{a}^\dagger(k_m)\right)^{n(k_m)}}{\sqrt{n(k_m)!}} \vert 0\rangle~.
\end{equation}
To measure how many particles of a certain momentum exist in the
field system, one can introduce the particle number operator
$\hat{N}(k)$,
\begin{equation}
\hat{N}(k)=\hat{a}^\dagger(k) \hat{a}(k)~.
\end{equation}
It is easy to check that
\begin{equation}
\hat{N}(k_i)\vert n(k_1),n(k_2),\cdots,n(k_m)\rangle=n(k_i)\vert n(k_1),n(k_2),\cdots,n(k_m)\rangle~.
\end{equation}
Thus, a field can be seen as a many-body system that consists of
free harmonic oscillators with different momentum. Properties of
the field system can be described exactly by the Weyl-Heisenberg
algebra,
\begin{equation}\label{algebra}
\begin{array}{l}
[\hat{a},\hat{a}^\dagger]=1~,\\[0.5cm]
[\hat{N},\hat{a}]=-\hat{a}~,~~~~~~[\hat{N},\hat{a}^\dagger]=\hat{a}^\dagger~,
\end{array}
\end{equation}
with the Fock space $\vert
n\rangle=\frac{(\hat{a}^\dagger)^n}{\sqrt{n!}}\vert 0\rangle$ as
its representation. All properties , in particular, the
thermodynamic properties of the many-body system can be
investigated systematically, by making use of the Hamiltonian
(\ref{hamiltonian}) and the Weyl-Heisenberg algebra
(\ref{algebra}).

The natural quantum of the quantum field theory on noncommutative
geometry is the so-called deformed harmonic oscillator. On the
topic of deformed oscillators, a lot of papers have been
published\cite{2}-\cite{b1}. Similar to the case of an ordinary
quantum field system, the many-body properties of a noncommutative
quantum field system can be described by a collection of the
deformed harmonic oscillators,
\begin{equation}
\hat{H}_q=\frac{1}{2}\sum_{\bf k}\hbar\omega_k\left(\hat{a}_q(k)\hat{a}_q^\dagger(k)+\hat{a}_q^\dagger(k)\hat{a}_q(k)\right)~.
\end{equation}
The deformed harmonic oscillator is related with the simple
harmonic oscillator as follows\cite{b2},
\begin{equation}
\hat{a}_q=\sqrt{\frac{[\hat{N}]_q}{\hat{N}}}\hat{a}~,~~~~~~~\hat{a}_q^\dagger=\hat{a}^\dagger
\sqrt{\frac{[\hat{N}]_q}{\hat{N}}}~,
\end{equation}
where we have used the notation $[x]_q=\displaystyle\frac{q^x-q^{-x}}{q-q^{-1}}$. \\
The deformed Weyl-Heisenberg algebra reads
\begin{equation}\label{1}
\begin{array}{l}
\hat{a}_q\hat{a}_q^{\dagger}-q\hat{a}_q^{\dagger}\hat{a}_q=q^{-\hat{N}} ~,\\[0.2cm]
[\hat{N},\hat{a}_q^{\dagger}]=\hat{a}_q^{\dagger}~,~~~~~~~
[\hat{N},\hat{a}_q]=-\hat{a}_q~.
\end{array}
\end{equation}
The representation of the deformed Weyl-Heisenberg algebra is
obtained by constructing the Fock space,
\begin{equation}
\begin{array}{l}
\vert n\rangle_q=\displaystyle\frac{(\hat{a}_q^\dagger)^n}{\sqrt{[n]_q!}}\vert 0\rangle~,\\[0.5cm]
\hat{a}_q \vert n\rangle_q=\sqrt{[n]_q}\vert n-1\rangle_q~,~~~~~~\hat{a}^\dagger_q \vert n\rangle_q=\sqrt{[n+1]_q}\vert n+
1\rangle_q~,
\end{array}
\end{equation}
where $[n]_q!\equiv [n]_q[n-1]_q\cdots [2]_q[1]_q$. It is easy to
check that one still can measure how many particles of a certain
momentum exist in the deformed field system by making use of the
particle number operator $\hat{N}(k)$,
\begin{equation}
\hat{N}(k_i)\vert n(k_1),n(k_2),\cdots,n(k_m)\rangle_q=n(k_i)\vert n(k_1),n(k_2),\cdots,n(k_m)\rangle_q~.
\end{equation}
The Hamiltonian of a deformed harmonic
oscillator\cite{b3}-\cite{b6} can be written as
\begin{equation}\label{4}
\hat{H}_q=\frac{\hbar\omega}{2}([\hat{N}]_q+[\hat{N}+1]_q)~.
\end{equation}
Below we will calculate the distribution function of a deformed
harmonic oscillator system by making use of the double-time Green
function technique.

\section{Distribution function}

The double-time Green function (retarded or advanced) for a Bose
system\cite{19,20} is generally defined as
\begin{equation}\label{6}
G_{\rho}(t-t')=-\frac{i}{2}\left((\rho+1)\theta(t-t')+(\rho-1)\theta(t'-t)\right){\langle}
[\hat{A}(t),\hat{B}(t')]{\rangle} ~,
\end{equation}
where we have used the notations,
$$\begin{array}{l}
\hat{A}(t)\equiv\exp(\frac{iHt}{\hbar})\hat{A}\exp(\frac{-iHt}{\hbar})~,\\[0.3cm]
{\langle}\hat{O}{\rangle}=Z^{-1}{\rm Tr}(e^{-{\beta} H}\hat{O})~, ~~~~~~~~
Z={\rm Tr}(e^{-{\beta}H})~,
\end{array}$$
here $\rho=1$ for the retarded one and $\rho=-1$ for the advanced one.

In Eq.(\ref{6}), setting
$$
\hat{A}(t)\equiv\hat{a}(t)~, ~~~~~~\hat{B}(t')\equiv\hat{a}^{\dagger}(t')~, ~~~~{\rho}=1~,
$$
(the retarded Green function is assumed), we get
\begin{equation}
\label{7}
G_R(t-t')=-i{\theta}(t-t'){\langle}\hat{a}(t)\hat{a}^{\dagger}(t')-\hat{a}^{\dagger}(t')\hat{a}(t){\rangle}~.
\end{equation}
The equation of motion of the retarded Green function  reads
\begin{equation}
\label{8} i\hbar\frac{\partial }{\partial
t}G_R(t-t')={\hbar}\left({\delta}(t-t'){\langle}\hat{a}(t)\hat{a}^{\dagger}(t')-\hat{a}^{\dagger}(t')\hat{a}(t){\rangle}+
{\theta}(t-t')\left\langle\frac{d{\hat{a}(t)}}{dt}\hat{a}^\dagger(t')-\hat{a}^\dagger(t')\frac{d\hat{a}(t)}{dt}\right\rangle
\right)~.
\end{equation}
The Heisenberg equation for the operator $\hat{a}(t)$
$$
\label{10}
i\hbar\frac{d{\hat{a}(t)}}{dt}=[\hat{a}(t),\hat{H}(t)]
$$
gives,
\begin{equation}
\label{11}
\frac{d\hat{a}(t)}{dt}=-i\frac{\omega}{2}\left(q^{\hat{N}+1}+q^{-(\hat{N}+1)}\right)\hat{a}(t)~.
\end{equation}
Then, we have
\begin{equation}
\label{12} i\hbar\frac{\partial }{\partial
t}G_{R}(t-t')={\hbar}{\delta}(t-t')+\frac{\hbar\omega}{2}\left(q^{{\langle}\hat{N}+1{\rangle}}+q^{-{\langle}\hat{N}+1{\rangle}}
\right)G_{R}(t-t')~.
\end{equation}
Making use of the Fourier transformation
\begin{equation}
G_R(t-t')=\int^\infty_{-\infty}\frac{dE}{2\pi\hbar}G_R(E)e^{-iE(t-t')/\hbar}~,
\end{equation}
we get
\begin{equation}
\label{13}
G_{R}(E)=\frac{{\hbar}}{E-\frac{\hbar\omega(q^{{\langle}\hat{N}+1{\rangle}}+q^{-{\langle}\hat{N}+1{\rangle}})}{2}}~.
\end{equation}
The spectrum theorem of Green functions tells us
\begin{equation}
\label{14}
{\langle}\hat{a}^\dagger(t')\hat{a}(t){\rangle}=i\int_{-\infty}^{\infty}dE\frac{(G_R(E+i0^{\dagger})-G_R(E-i0^{\dagger}))
e^{-iE(t-t')/\hbar}}{2{\pi}{\hbar}(e^{{\beta}E}-1)}~.
\end{equation}
Then, we can write formally the distribution function of a
deformed Bose gas as
\begin{equation}
\label{15}
n(\varepsilon)={\langle}\hat{N}{\rangle}=\frac{1}{\exp(\frac{\hbar\omega\beta(q^{{\langle}\hat{N}+1{\rangle}}+q^{-{\langle}
\hat{N}+1{\rangle}})}{2})-1}~.
\end{equation}
Evidently, in the case of ${q\rightarrow}1$, $n(\varepsilon)$
recovers the  familiar Bose-Einstein distribution function. In
principle, all interesting properties of a deformed Bose gas can
be obtained by solving this equation for distribution function.
This is to be one of the most advantageous features of the Green
function approach.

Now we try to solve Eq.(\ref{15}) to gain an insight into the
influence of the deformation on the distribution function and to
compare the deformed distribution function with the Bose-Einstein
distribution function.

Expanding the right hand side of Eq.(\ref{15}) in the neighborhood
of $q=1$ to the second order, we get an exactly solvable equation
for the deformed distribution function. The solution is of the
form,
\begin{equation}\label{016}
\begin{array}{rcl}
n(\varepsilon)&=&\displaystyle-\frac{1}{\delta^2 \beta\varepsilon e^{\beta\varepsilon}}\left(\delta^2 \beta\varepsilon
e^{\beta\varepsilon}
+\left(e^{\beta\varepsilon}-1\right)^2\right)\\[0.5cm]
& &+\displaystyle\frac{1}{\delta^2 \beta\varepsilon e^{\beta\varepsilon}}\sqrt{\left(\delta^2\beta\varepsilon e^{\beta\varepsilon}+
\left(e^{\beta\varepsilon}-1\right)^2\right)^2-\delta^2\beta\varepsilon e^{\beta\varepsilon}\left(\delta^2\beta\varepsilon
e^{\beta\varepsilon}
-2(e^{\beta\varepsilon}-1)\right)}~,
\end{array}
\end{equation}
where we have used the notation $\delta\equiv q-1$ for short.

We note that it is difficult to compare the distribution function
of the deformed Bose gas, Eq.(\ref{016}), with the Bose-Einstein
distribution function directly. For the aim, we rewrite
Eq.(\ref{016}) as follows,
\begin{equation}
n(\varepsilon)=\frac{(e^{\beta\varepsilon}-1)-2^{-1}\delta^2{\beta\varepsilon} e^{\beta\varepsilon}}{(e^{\beta\varepsilon}-1)^2+
\delta^2{\beta\varepsilon}
e^{\beta\varepsilon}}-\frac{\delta^2{\beta\varepsilon}e^{\beta\varepsilon}\left(\delta^2{\beta\varepsilon}
e^{\beta\varepsilon}-2\left(e^{\beta\varepsilon}-1\right)\right)^2}{\left(2\delta^2{\beta\varepsilon}e^{\beta\varepsilon}
+2\left(e^{\beta\varepsilon}-1\right)^2\right)^3}~.
\end{equation}
One can further expand the distribution function into a series
over deformation parameter $\delta$,
\begin{equation}
n({\varepsilon})=\frac{1}{e^{{\beta}{\varepsilon}}-1}-\delta^2\left(\frac{{\beta}{\varepsilon}e^{{\beta}{\varepsilon}}}
{2(e^{{\beta}{\varepsilon}}-1)^2}+\frac{{\beta}{\varepsilon}e^{{\beta}{\varepsilon}}}{(e^{{\beta}{\varepsilon}}-1)^3}
+\frac{{\beta}{\varepsilon}e^{{\beta}{\varepsilon}}}{2(e^{{\beta}{\varepsilon}}-1)^4}\right)~.
\end{equation}
From the above form of the distribution function of the deformed
Bose gas, we see clearly that, at the case of $q=1$, the
Bose-Einstein distribution is recovered. To further investigate
the effect of deformation on the distribution function, we present
a plot of the distribution function in Fig.1.
\begin{center}
  \includegraphics{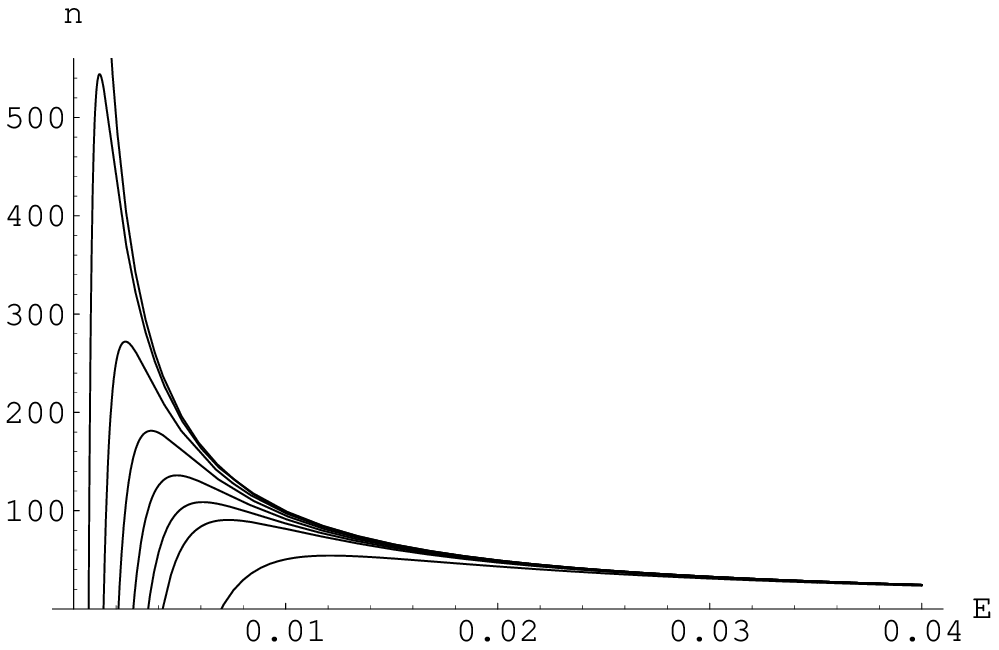}

{\bf FIG.1}~~~~The distribution function of a deformed Bose gas\\
{\scriptsize ( start from up $q=1.000,~ 1.001,~ 1.002,~ 1.003,~
1.004,~ 1.005,~ 1.006, ~1.010$, here we have used the notation
${\beta}{\varepsilon}=E$.)}
\end{center}
To describe behaviors of the distribution function of the deformed
Bose gas near the ground state more accurately, we give another
plot of the deformed distribution function in Fig.2.

\begin{center}
\includegraphics{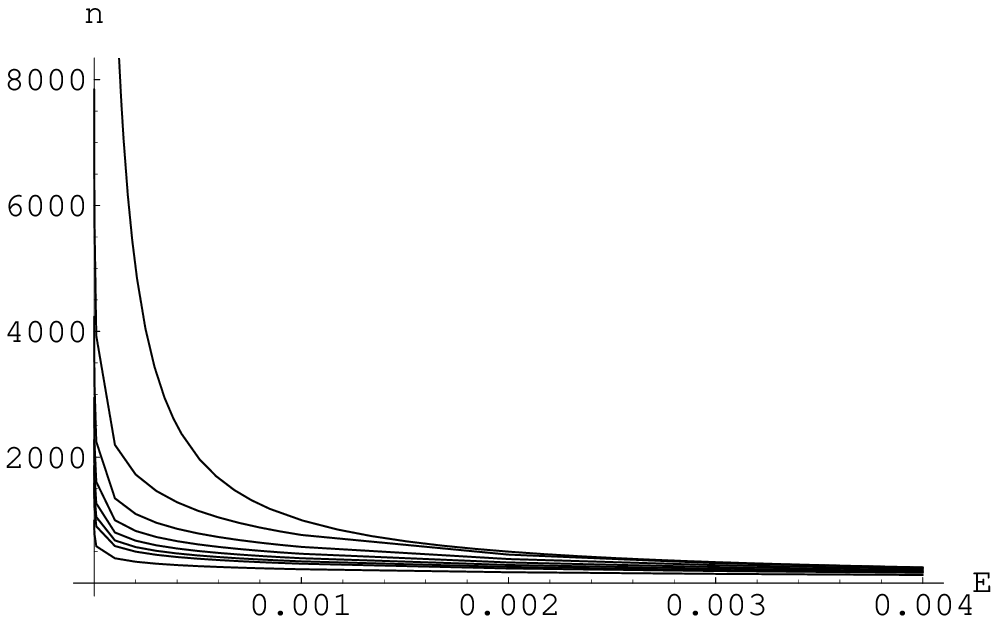}

{\bf FIG. 2}~~~~The behaviors of distribution function\\
~~~~~~~~of a deformed Bose gas near the ground state\\
{\scriptsize ~~~~~~~(start from up $q=1.000,~ 1.001,~ 1.002,~
1.003,~ 1.004,~ 1.005,~ 1.006, ~1.010$, here we have used the
notation ${\beta}{\varepsilon}=E$.)}
\end{center}

\section{Thermodynamic properties}

In this section, we concentrate on the study of thermodynamic
properties of a deformed photonic field. As being well-known, the
dispersion relation of a photon is
\begin{equation}
\epsilon=cp~.
\end{equation}
We know that a photonic field can be described by a collection of
simple harmonic oscillators and have relation
\begin{equation}
\hat{H}=\hbar\omega\left(\hat{N}+\frac{1}{2}\right)~.
\end{equation}
One can then read out that
\begin{equation}
p=\frac{\hbar\omega}{c}\left(n+\frac{1}{2}\right)~.
\end{equation}
This relation and the Hamiltonian (\ref{4}) give the dispersion
relation of a deformed Bose gas,
\begin{equation}
\label{16}
\varepsilon=\frac{1}{2}\hbar\omega\left(\left[\frac{c}{\hbar\omega}p+\frac{1}{2}\right]_q+
\left[\frac{c}{\hbar\omega}p-\frac{1}{2}\right]_q\right)~.
\end{equation}

Expanding the right hand side of (\ref{16}) in the neighborhood of
$q=1$ to the second order, we get
\begin{equation}
\varepsilon=cp-\frac{\delta^2}{24}cp+\frac{\delta^2}{6{\hbar}^2{\omega}^2}c^3p^3
~.
\end{equation}
The density of energy levels is as follows,
\begin{equation}
\begin{array}{rcl}
\label{17}
a(\varepsilon)&=&\displaystyle\frac{8{\pi}V}{3h^3c^3}\cdot\frac{3\sqrt{3B}\varepsilon+\sqrt{4A^3+27B{\varepsilon}^2}}{4B\sqrt{4A^3+27B
{\varepsilon}^2}}\cdot\frac{\left(-2\times6^{1/3}A+\left(18\sqrt{B}{\varepsilon}+2\sqrt{3(4A^3+27B{\varepsilon}^2)}
\right)^{2/3}
\right)^2}{\left(18\sqrt{B}{\varepsilon}+2\sqrt{3(4A^3+27B{\varepsilon}^2)}\right)^{1/3}}\\[1cm]
& &\times\displaystyle
\frac{9\sqrt{B}\varepsilon+\sqrt{3(4A^3+27B{\varepsilon}^2)}+6^{1/3}A\left(18\sqrt{B}{\varepsilon}+2\sqrt{3(4A^3+27B
{\varepsilon}^2)}\right)^{1/3}}{
\left(9\sqrt{B}\varepsilon+\sqrt{3(4A^3+27B\varepsilon^2)}\right)^{2}} ~,
\end{array}
\end{equation}
where we have used the notations
$A\equiv\displaystyle(1-\frac{\delta^2}{24})c$ and $B\equiv\displaystyle\frac{\delta^2}{6{\hbar}^2{\omega}^2}c^3$.

The mean value of the total energy of the system has the form,
\begin{equation}
\label{19}
\bar{E}=\int_0^{\infty}{\varepsilon}n({\varepsilon})a({\varepsilon})d{\varepsilon}~.
\end{equation}
To calculate $\bar{E}$, we expand the right hand side of
(\ref{17}) to the third order in the neighborhood of $q=1$ and
rewrite the density of energy levels as the form
\begin{equation}
a({\varepsilon})=\frac{8{\pi}V}{3h^3c^3}\left(3-\frac{17}{8}\delta^2\right){\varepsilon}^2 ~.
\end{equation}
Then we get
\begin{equation}
\bar{E}=\frac{8{\pi}^5
V}{15}\frac{(kT)^4}{(hc)^3}-{\pi}V\delta^2\left(\frac{8{\pi}^2}{3}+\frac{11{\pi}^4}{15}+48{\zeta}(3)\right)\frac{(kT)^4}
{(hc)^3}~,
\end{equation}
 where ${\zeta}(n)$ is the Riemann zeta function.

We can further calculate the grand partition function $\ln\Xi$
through the thermodynamic relation $\bar{E}=-\frac{\partial
}{\partial \beta}\ln\Xi$ ,
\begin{equation}
 {\ln}{\Xi}=\frac{8{\pi}^5
V}{45h^3c^3}(kT)^3-\frac{{\pi}V\delta^2}{h^3c^3}\left(\frac{8{\pi}^2}{9}+\frac{11{\pi}^4}{45}+16{\zeta}(3)\right)(kT)^3 ~.
\end{equation}
It is a straightforward calculation to obtain thermodynamic
functions by making use of the grand partition function. The
pressure of the deformed Bose gas is
\begin{equation}\label{pressure}
P=\frac{1}{\beta}\frac{\partial}{\partial V
}{\ln}{\Xi}=\frac{8{\pi}^5}{45h^3c^3}(kT)^4-\frac{{\pi}\delta^2}{h^3c^3}\left(\frac{8{\pi}^2}{9}+\frac{11{\pi}^4}{45}+
16{\zeta}(3)\right)(kT)^4~.
\end{equation}
The free energy of the deformed Bose gas is of the form,
\begin{equation}
F=-\frac{8{\pi}^5V}{45}\frac{(kT)^4}{(hc)^3}+{\pi}V\delta^2\left(\frac{8{\pi}^2}{9}+\frac{11{\pi}^4}{45}+16{\zeta}(3)\right)
\frac{(kT)^4}{(hc)^3}~.
\end{equation}
 The entropy of the deformed Bose gas is as
\begin{equation}
S=\frac{U-F}{T}=\frac{32{\pi}^5}{45}\frac{k^4T^3
}{h^3c^3}V-{\pi}\delta^2\left(\frac{32{\pi}^2}{9}+\frac{44{\pi}^4}{45}+64{\zeta}(3)\right)\frac{k^4T^3}{h^3c^3}V~.
\end{equation}
The first term in the above equation is just the entropy of an
ordinary Bose gas. And the terms following it denote the effect of
noncommutative deformation of geometry on the entropy. It is clear
that the effect can not be ignored at least in the high
temperature region, where the early universe lived. We give a plot
of the entropy vs temperature for different deformation parameters
in Fig. 3.
\begin{center}
\includegraphics{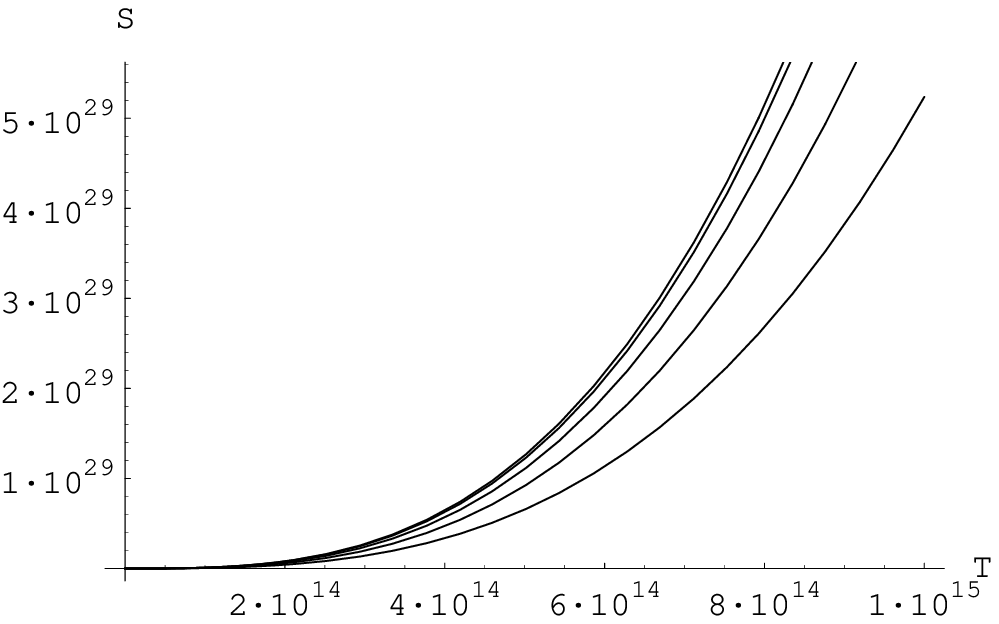}

{\bf FIG.3}~~~~The entropy per unit volume of a deformed Bose gas\\
{\scriptsize ( start from up $q=1.0,~ 1.1,~1.2,~ 1.3,~ 1.4$).}
\end{center}
Finally, the specific heat of the deformed Bose gas is
\begin{equation}
C_V=T(\frac{\partial S}{\partial
T})_V=\frac{32{\pi}^5
V}{15}\frac{k^4T^3}{h^3c^3}-{\pi}V\delta^2\left(\frac{32{\pi}^2}{3}+\frac{44{\pi}^4}{15}+192{\zeta}(3)\right)\frac{k^4T^3}
{h^3c^3}~.
\end{equation}
At the case of $q=1$, we recover the result for an ordinary Bose
gas. To show the effect of deformation on the specific heat, we
give a plot of the specific heat per unit volume of a deformed
Bose gas vs temperature for different deformation parameters in
Fig. 4.

From the above discusses, we can also obtain the {\em Planck's
formula} for the deformed Bose gas,

\begin{center}
\includegraphics{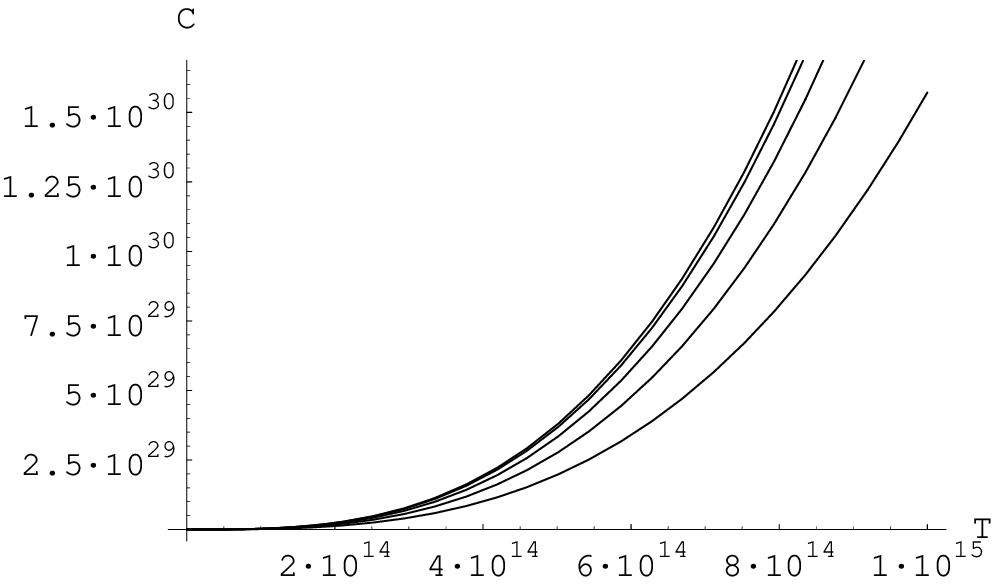}

{\bf FIG.4}~~~~The specific heat per unit volume of a deformed Bose gas\\
{\scriptsize ( start from up $q=1.0,~ 1.1,~1.2,~ 1.3,~ 1.4$).}

\end{center}
\begin{equation}
\begin{array}{rcl}
u(\omega,T)&=&\displaystyle\frac{{\hbar}{\omega}^3}{c^3{\pi}^2(e^{{\beta}{\hbar}{\omega}}-1)}-\frac{17\delta^2{\hbar}{\omega}^3}
{24c^3{\pi}^2(e^{{\beta}{\hbar}{\omega}}-1)}-\frac{\delta^2{\beta}{\hbar}^2{\omega}^4e^{{\beta}{\hbar}{\omega}}}{2c^3{\pi}^2
(e^{{\beta}{\hbar}{\omega}}-1)^2}\\[0.5cm]
& & \displaystyle
-\frac{\delta^2{\beta}{\hbar}^2{\omega}^4e^{{\beta}{\hbar}{\omega}}}{c^3{\pi}^2(e^{{\beta}{\hbar}{\omega}}-1)^3}-
\frac{\delta^2{\beta}{\hbar}^2{\omega}^4e^{{\beta}{\hbar}{\omega}}}{2c^3{\pi}^2(e^{{\beta}{\hbar}{\omega}}-1)^4}~.
\end{array}
\end{equation}
\begin{center}
\includegraphics{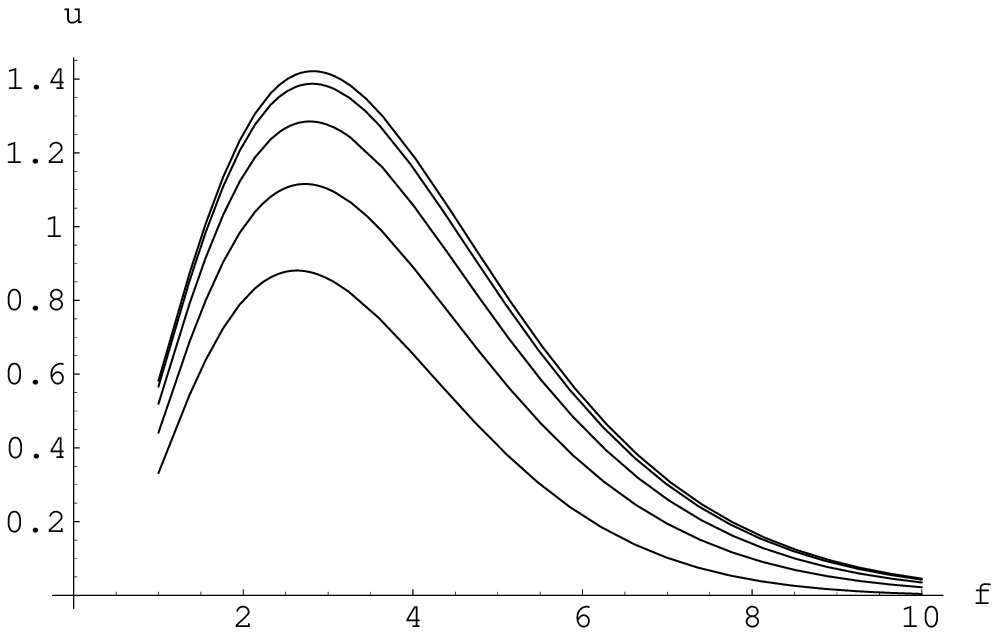}

{\bf FIG.5}~~~~The {\em Planck's formula} of a deformed Bose gas\\
{\scriptsize ( start from up $q=1.0,~ 1.1,~1.2,~ 1.3,~ 1.4$, here
we have used the notation $f={\beta}{\hbar}{\omega}$.})
\end{center}
In Fig.5, we give a plot of the {\em Planck's formula} for the
deformed Bose gas. The first term in equation(46) describes an
ideal gas consisting of photons. The linearity of the equations of
electrodynamics on commutative spacetime guarantees that photons
do not interact with one another. However, the other terms in this
equation denote interactions among photons or nonlinearity of
equations of electrodynamics caused by noncommutative deformation
of spacetime.
At high frequencies ($\beta\hbar\omega\gg 1$), we
have
\begin{equation}
\begin{array}{rcl}
u(\omega,T)&=&\displaystyle\frac{{\hbar}{\omega}^3e^{-{\beta}{\hbar}{\omega}}}{c^3{\pi}^2}-\frac{17\delta^2{\hbar}
{\omega}^3e^{-{\beta}{\hbar}{\omega}}}{24c^3{\pi}^2}-\frac{\delta^2{\beta}{\hbar}^2{\omega}^4e^{-{\beta}{\hbar}{\omega}}}
{2c^3{\pi}^2}\\[0.5cm]
& &\displaystyle
-\frac{\delta^2{\beta}{\hbar}^2{\omega}^4e^{-2{\beta}{\hbar}{\omega}}}{c^3{\pi}^2}-\frac{\delta^2{\beta}{\hbar}^2{\omega}^4
e^{-3{\beta}{\hbar}{\omega}}}{2c^3{\pi}^2}~.
\end{array}
\end{equation}
This is the {\em Wien's formula} of the deformed Bose gas. We give
a plot of the {\em Wien's formula} for different deformation
parameters in Fig.6. In spite of a complex form of the {\em
Planck's formula} and the {\em Wien's formula}, we note that the
density of the spectral frequency distribution of the energy of
the deformed Bose gas has a maximum at a frequency ${\omega}_m$
(${\beta}{\hbar}{\omega}_m{\approx}2.8$), which is roughly
independent of deformation parameter. Thus the {\em displacement
law} is still remained for the deformed Bose gas.

\begin{center}
\includegraphics{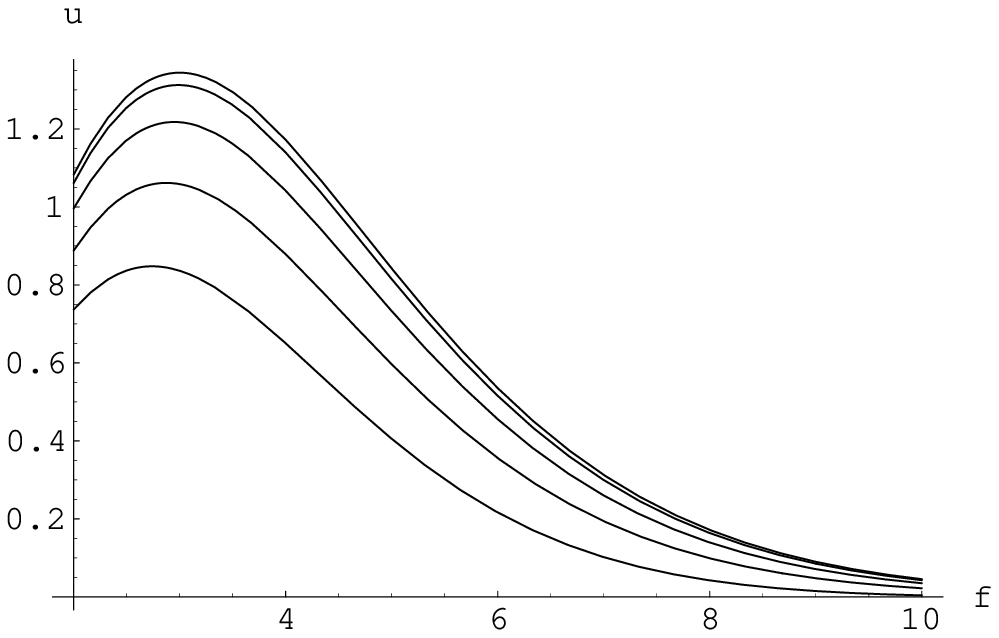}

{\bf FIG.6}~~~~The {\em Wien's formula} of a deformed Bose gas\\
{\scriptsize ( start from up $q=1.0,~ 1.1,~1.2,~ 1.3,~ 1.4$, here
we have used the notation $f={\beta}{\hbar}{\omega}$).}
\end{center}
In the opposite limiting case of low frequencies ($\beta\hbar\omega\ll 1$), we get the {\em Rayleigh-Jeans formula}
for the deformed Bose gas,
\begin{equation}
\begin{array}{rcl}
 u(\omega,T)&=&\displaystyle\frac{kT{\omega}^2}{c^3{\pi}^2}-\frac{17\delta^2kT{\omega}^2}{24c^3{\pi}^2}-\frac{\delta^2
kT{\omega}^2}{2c^3{\pi}^2}(\frac{{\hbar}{\omega}}{kT}+1)\\[0.5cm]
& &\displaystyle
-\frac{\delta^2(kT)^2{\omega}}{c^3{\pi}^2{\hbar}}(\frac{{\hbar}{\omega}}{kT}+1)-\frac{\delta^2(kT)^3}{2c^3{\pi}^2{\hbar}^2}
\left(\frac{{\hbar}{\omega}}{kT}+1\right)~.
\end{array}
\end{equation}
Fig.7 shows a graph corresponding to the {\em Rayleigh-Jeans
formula} with different deformation parameters.

\begin{center}
\includegraphics{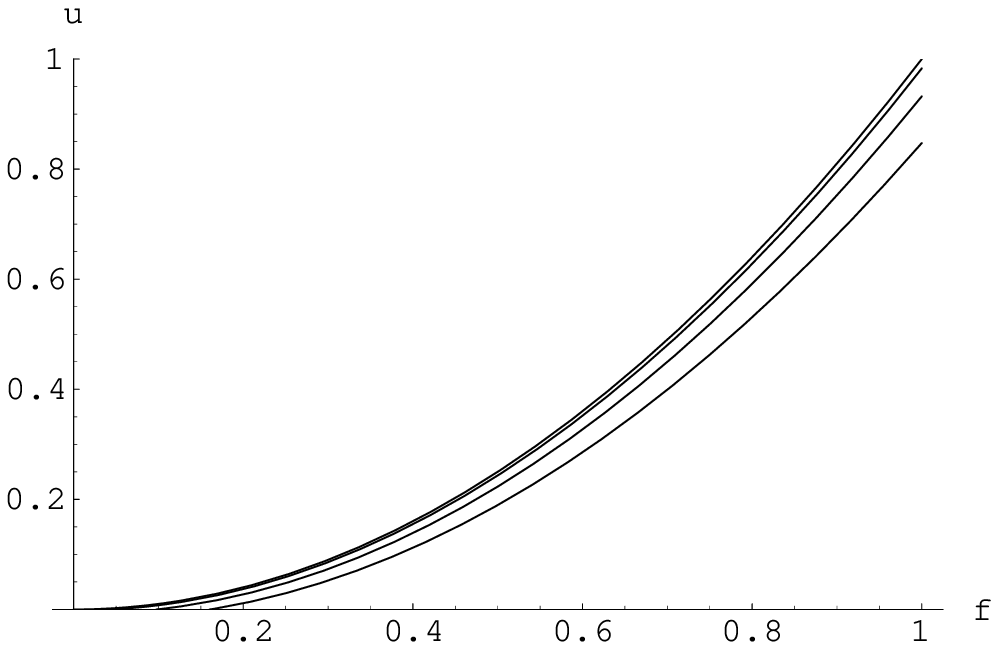}

{\bf FIG.7}~~~~The {\em Rayleigh-Jeans formula} of a deformed Bose gas\\
{\scriptsize ( start from up $q=1.00,~ 1.06,~1.12,~ 1.18$).}
\end{center}

We obtain the {\em Stefan-Boltzmann's law} for the deformed Bose
gas of the form,
\begin{equation}
J=\frac{{\pi}^2k^4
}{60{\hbar}^3c^2}T^4-\frac{\delta^2k^4}{32{\pi}^2{\hbar}^3c^2}\left(\frac{8{\pi}^2}{3}+\frac{11{\pi}^4}{15}+
48{\zeta}(3)\right)T^4~.
\end{equation}
\begin{center}
\includegraphics{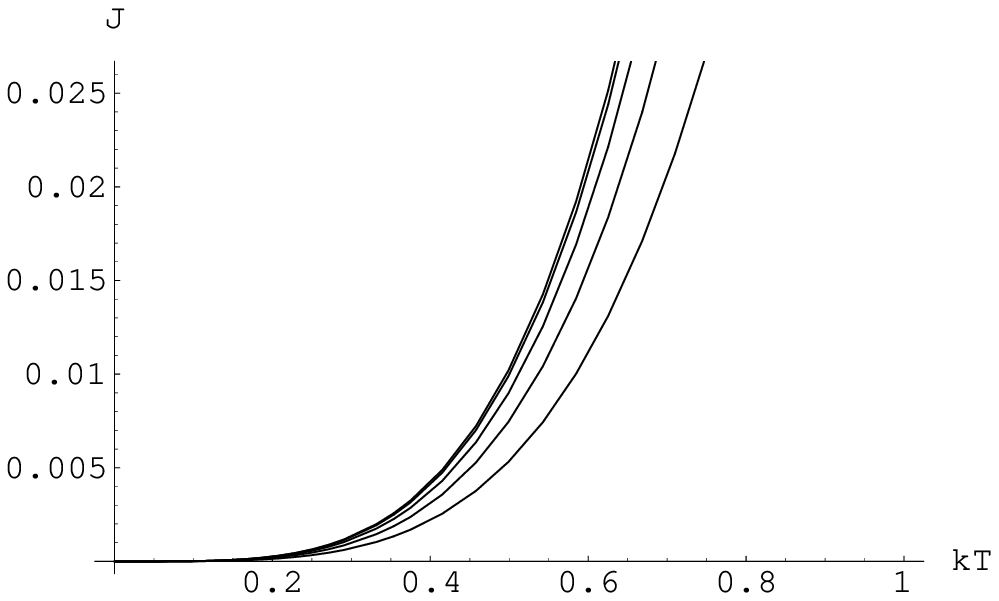}

{\bf FIG.8}~~~~The {\em Stefan-Boltzmann law} of a deformed Bose gas\\
{\scriptsize ( start from up $q=1.0,~ 1.1,~1.2,~ 1.3,~ 1.4$).}

\end{center}
Similar to the case of an ideal Bose gas, the total energy of the
deformed Bose gas is proportional to the fourth power of the
temperature. But the {\em Stefan-Boltzmann's constant} is
effectively reduced by the noncommutative deformation of geometry.
The fact gives also a signal that the deformed Bose gas is a
strongly correlated system. We give a plot of the {\em
Stefan-Boltzmann's law} for different deformation parameters in
Fig.8.

In an adiabatic expansion (or compression) process, the entropy of the system is a constant and the valume and temperature are related by
$VT^3=$constant. From (\ref{pressure}), we can get the equation of state,
\begin{equation}
PV^{4/3}=c_q~.
\end{equation}

\section{Concluding Remarks}

The noncommutative geometry plays a very important role in the
study of the Planck scale  physics. Quantum field theory and
superstring theory (or M theory) on noncommutative spacetime have
been studied for a long time. It is well-known that the most
realistic laboratory for testing the Planck scale physics is the
early universe. The very early universe was opaque due to constant
interchange of energy between matter and radiation. It is clear
that the radiation field at the early universe can not be
described well by an ideal Bose gas system.  There should be
strong correlations among photons at this stage of evolution of
the universe. In fact, attention has been paid recently to the
investigation on the deformed radiation dominated universe.
Preliminary results show that spacetime quantization gives rise to
inflation as the radiation temperature exceeds the Planck
temperature. Of course, the study in this direction is only on its
very beginning. Many efforts should be made to set up a
self-contained inflation theory on noncommutative spacetime. We
still have little knowledge on thermodynamics of the deformed
radiation field.

From the linearity of the equations of electrodynamics on
commutative spacetime, we know that photons do not interact with
one another. The thermodynamics of the photonic field can be
described well by an ideal Bose gas system.  However, the photons
in the deformed radiation field have strong correlations among
themselves. The strong correlations among the photons are caused
by the nonlinearity of equations of electrodynamics on
noncommutative spacetime. It is a difficult task to deal with the
interactions among the photons. Numerous papers have been
published\cite{5}-\cite{18} on the thermodynamics of the deformed
photonic field. However, we still have not got a satisfactory
distribution function of the deformed Bose gas.

In this paper, we have treated the deformed photonic field as a
strongly correlated system by making use of the Green's function
theory. Equations of motion of the Green's function for the
deformed photonic field was set up. The Tyablikov
procedure\cite{21} was used to decouple these equations of motion.
A general form of statistical distribution function for the
deformed photonic field was got systematically. Based on this
distribution function, we calculated the free energy, the entropy
and the specific heat of the deformed photonic field in detail.
The equation of state was obtained. We found that the {\em
Planck's formula} for the deformed Bose gas is really different
from the ordinary one. We got terms in {\em Planck's formula},
which correspond to the interactions among photons. Similar to the
case of an ideal Bose gas, the total energy of the deformed Bose
gas is proportional to the fourth power of the temperature. But
the {\em Stefan-Boltzmann's constant} is effectively reduced by
the noncommutative deformation of geometry. We noted that the
density of the spectral frequency distribution of the energy of
the deformed Bose gas has a maximum at a frequency ${\omega}_m$
(${\beta}{\hbar}{\omega}_m{\approx}2.8$), which is roughly
independent of deformation parameter. Thus the {\em displacement
law} is still remained for the deformed Bose gas. The results
obtained in the paper can be used to investigate properties of the
early universe ({\em e.g.} inflation model on noncommutative
spacetime). All of these works are in progress.

\bigskip\bigskip

\centerline{\large\bf Acknowledgements}

The work was supported in part by the Natural Science Foundation
of China.

\vspace{1cm}

\

\end{document}